\documentclass[review,12pt]{elsarticle}

\usepackage{amssymb}
\usepackage{longtable}
\usepackage{bm}
\usepackage{multirow}
\usepackage{hyperref}
\usepackage{ulem}
\hypersetup{colorlinks,citecolor=blue,filecolor=blue,linkcolor=blue,urlcolor=blue}
\usepackage{color}
\usepackage{dcolumn}
\usepackage{tabularx}
\usepackage{amsmath}
\usepackage[usenames,dvipsnames]{xcolor}

\definecolor{awesome}{rgb}{1.0, 0.13, 0.32}
\newcommand{\iso}[2]{$^{#1}$#2}

\journal{}

\begin{document}

\begin{frontmatter}


\title{Kinematics Calibration and Excitation Energy Reconstruction for Solenoidal Spectrometers}


\author[inst1,inst2]{T.~L.~Tang}
\ead{ttang@anl.gov}
\address[inst1]{Department of Physics, Florida State University, Tallahassee, Florida 32306, USA}
\address[inst2]{Physics Division, Argonne National Laboratory, Lemont, IL 60439, USA}

\begin{abstract}

This work introduces a novel method for reconstructing excitation energy ($E_x$) and center-of-mass scattering angle ($ \theta_{cm}$) from energy-position data in solenoidal spectrometers, addressing challenges posed by non-linearities at small scattering angles. The approach employs a robust calibration of experimental energy-position data using known excited states, followed by an analytical inverse transformation based on relativistic kinematics and cyclotron motion. Integrated into the HELIOS online analysis routines, this method enables real-time generation of excitation energy spectra and angular distributions during experiments, improving efficiency and accuracy over traditional projection-based methods. The method’s effectiveness is demonstrated using the $^{25}$Mg$(d, p)$ reaction, highlighting its ability to handle forward-angle data and produce precise kinematic reconstructions.

%Solenoidal spectrometers are gaining popularity worldwide due to their large acceptance and superior Q-value resolution in reactions involving radioactive beams. This work introduces a calibration method for the light-recoil energy that leverages known excited states. Building on this calibration, we describe an algorithm that transforms energy-position data into excitation energy and center-of-mass scattering angle. This approach has been integrated into the HELIOS online analysis routines, enabling the fast deduction of accurate excitation energy spectra and angular distributions during experiments, which provides informative decision-making.

\end{abstract}


\end{frontmatter}

%%\linenumbers

\section{Introduction}

Solenoidal spectrometers~\cite{Wuosmaa2007} are specialized instruments used in nuclear reaction studies of the type A($a$,$b$)B, where only the residual nucleus B may be excited. These spectrometers relate the excitation energy of particle B ($E_x$, in MeV) and the center-of-mass scattering angle ($\theta_{cm}$) to the kinetic energy ($E$, in MeV) of the light recoil charged particle $b$ and its cyclotron position $Z_0$ (in mm), corresponding to the axial position where particle $b$ crosses the beam axis after completing one cyclotron period. This transformation is denoted as $(E_x, \theta_{cm}) \rightarrow (E, Z_0)$. Due to the presence of the magnetic field, the relativistic relationship between $E$ and $Z_0$ is given by~\cite{Tang2023}:
\begin{equation}\label{eq1}
E + m = \frac{E_{cm}}{\gamma} + \alpha \beta Z_0, \quad \alpha = \frac{q c B}{2\pi}, \quad \gamma^2 = \frac{1}{1 - \beta^2},
\end{equation}
Here, $m$, $q$, $E_{cm}$ are the mass (in MeV/c$^2$), the charge state, and the total energy (in MeV)) in the center-of-mass frame of the light-recoil $b$, respectively. $\gamma$ and $\beta$ are the Lorentz boost factors from the laboratory frame to the center-of-mass frame. The constant $c$ represents the speed of light with a value approximately $300$~ mm/ns, and $B$ is the magnetic field in Tesla. The excitation energy $E_x$ is implicitly included in $E_{cm}$, with higher excitation energies corresponding to smaller $E_{cm}$ values. Note that $E$ is linear to $Z_0$. 

An axial detector array, positioned in the center of the beam, measures the energy (denoted $e$, in channel) of the light recoil particle $b$ and its position z. A conventional approach is to first convert the measured energy $e$ to the calibrated energy $E$ (in MeV), then transform the energy $E$ and position $Z_0$ into $E_x$ by projecting the excitation lines for each detector, after that an energy spectrum for each detector is aligned and calibrated by adjusting the offset and scale to match the known states. The corresponding $\theta_{cm}$ values are deduced from the $Z_0$ positions using kinematics simulations. %When only one state is known and can be compared, scaling to produce the correct energy spectrum is difficult.

\begin{figure}[ht!]
\centering
%trim= left bottom right top
\includegraphics[trim=0cm 0cm 0cm 0cm, clip, width=12cm]{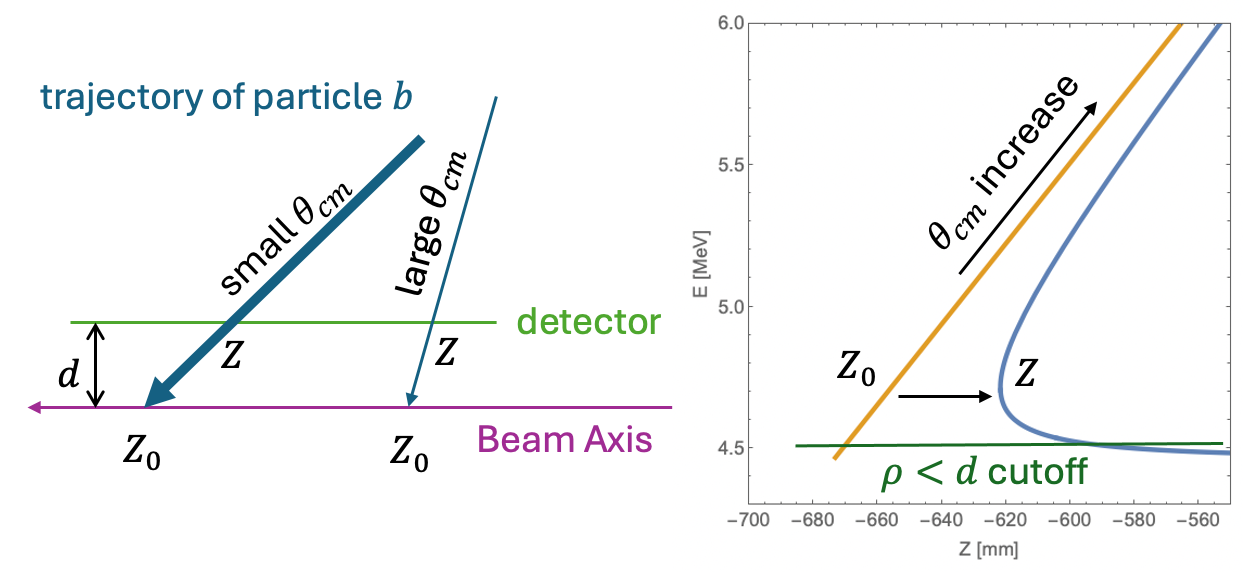}
\caption{\label{fig:bending}. The effect of finite detector size on the $Z$ position. This lead to bending of the $E-Z$ curve of Eq.~\ref{eq1}. See main text for detail.}
\end{figure}

However, due to the finite size of the axial detector array, the measured position ($Z$) differs from the true cyclotron crossing position ($Z_0$). The relationship is given by:
\begin{equation}\label{eq2}
Z = Z_0 \left( 1 - \frac{1}{2\pi} \sin^{-1}\left(\frac{d}{\rho}\right) \right) \approx Z_0 \left( 1 - \frac{1}{2\pi} \frac{d}{\rho} \right), \quad \text{for}~d \ll \rho
\end{equation}
where $d$ is the perpendicular distance between the axial detector surface and the beam axis, and $\rho \propto \sin\theta_{cm}$ is the radius of the cyclotron motion. The effect of this positional shift is illustrated in the left side of Fig.~\ref{fig:bending}. When particle $b$ strikes the detector at a shallow incidence angle (the thicker line), corresponding to a smaller $\theta_{cm}$ and $\rho$, the detected position $Z$ deviates more significantly from $Z_0$. In cases where $\theta_{cm}$ or $\rho$ is very small such that $\rho < d$, the particle $b$ cannot reach the detector and will not be detected. In the $E-Z$ plot, the straight line predicted by Eq.~\ref{eq1} (orange) becomes a bent curve described by Eq.~\ref{eq2} (blue), as shown on the right side of Fig.~\ref{fig:bending}.

%----------------------FIGURE 1-----------------
%----------------------------------------------

The approximation in Eq.~\ref{eq2} holds for $d/\rho < 0.2$. 
%This correction causes bending in the linear relationship from Eq.\ref{eq1}, as shown in Fig.10 of Ref.\cite{Wuosmaa2007}. 
For large $\theta_{cm}$ values, $\rho$ is also large, and $d/\rho \approx 0$, then $Z \approx Z_0$, leaving the $E$-$Z_0$ relationship largely unaffected. However, for $\theta_{cm} < 20^\circ$ (depending on $d/\rho$), the bending becomes significant, making the projection method less effective and difficult. The difficulty is due to the fact that the curvatures of the $E-Z$ curve for different states are slightly different in a single detector, which has a fixed position coverage. This means the $\theta_{cm}$ coverage differs for each excited state. As a result, the projection method cannot achieve optimal energy resolution across all states. Also, the projection method cannot deduce the $\theta_{cm}$. To eliminate the contribution from small $\theta_{cm}$ in the $E_x$ spectrum, a manual graphical gate must be applied. This introduces a systematic error in the yield, as the $\theta_{cm}$ coverage could vary between different excited states.

%At forward angles ($\theta_{cm} < 8^\circ-10^\circ$ in HELIOS~\cite{Lightall2010}), the $E$-$Z$ curves for different states overlap, leading to degeneracies where distinct $(E_x, \theta_{cm})$ combinations result in the same $(E, Z)$. Consequently, forward angles become unusable in most cases.

We present a new method to systematically and efficiently extract excitation energy from the bent $E-Z$ curve. This approach involves first calibrating the measured energy $e$ using reaction kinematics, followed by applying an inverse transformation, $(E, Z) \rightarrow (E_x, \theta_{\text{cm}})$, to directly obtain the excitation energy and center-of-mass scattering angle at once.

%######################################################
\section{Kinematics Calibration}

For reactions where several excited states are well known, the $E$-$Z$ curves corresponding to these known states can be calculated using Eqs.~\ref{eq1} and~\ref{eq2}. Let us denote these theoretical kinematics curves as $E = f_i(Z)$. The goal is to scale ($a$) and offset ($b$) the measured experimental energy $e$ (in channel) to match the theoretical energy $E$ (in MeV). The position of the array is assumed to be accurately measured.

To achieve this, we use a minimum chi-squared method to determine the calibration parameters $(a, b)$. Each data point $(e_j, z_j)$ is assumed to originate from a specific excited state $i$, for which there is a single theoretical curve $f_i$ that provides the best fit. The energy $e_j$ is in channel and the $z$-position ($z_j$) is assumed to be known correctly. The squared distance between the scaled and offset experimental energy and the theoretical energy is given by:
\begin{equation}
d^2_{ij}(a, b) = \left( a e_j + b - f_i(z_j) \right)^2.
\end{equation}

The calibration parameters $(a, b)$ are obtained by minimizing the sum of squared distances for all data points:
\begin{equation}
\chi^2 = \sum_{ij} \min\left( d^2_{ij}(a, b), \tau\right).
\end{equation}

A threshold ($\tau$) is introduced to exclude contributions from noise or unknown states. Data points for which the distance $|d_{ij}|$ exceeds this threshold are not counted as valid. The minimization of $\chi^2$ implicitly maximize the number of valid data points ($N$) that satisfy the condition $|d_{ij}| < \tau$. 

This method ensures accurate calibration of the experimental energy by leveraging the theoretical curves of known states. Even at small center-of-mass scattering angles, where $E$-$Z$ curves may overlap and bending becomes significant, this approach maintains reliability by identifying the best-fitting theoretical curve for each data point.

%######################################################
\section{Reconstruction of $(E_x,\theta_{cm})$}\label{sec_3}

After calibrating the measured energy $e$ in channel to the calibrated energy $E$ in MeV for the light recoil, we proceed to transform $(E,Z)$ to $(E_x,\theta_{cm})$. By combining Eqs.~\ref{eq1} and \ref{eq2}, we obtain the following relationship:
\begin{equation}\label{eq_3}
\alpha \beta \gamma Z = (\gamma y - E_{cm}) \left( 1 - \frac{1}{2\pi}\frac{d}{\rho} \right), \quad y = E + m,
\end{equation}
where $E_{cm}$ and $\rho$ are given by:
\begin{equation}\label{eq_4}
E_{cm} = \sqrt{m^2 + k^2}, \quad \rho = \frac{k \sin\theta_{cm}}{2\pi \alpha},
\end{equation}
with $k$ being the momentum of particle $b$ in the center-of-mass frame. The energy of particle $b$ in the lab's frame is expressed as:
\begin{equation}\label{eq5}
y = \gamma \sqrt{m^2 + k^2 } - \gamma \beta k \cos\theta_{cm}.
\end{equation}
Substituting $k = m \tan(x)$, where $0 < x < \pi/2$, and eliminating $\theta_{cm}$ and $\rho$ using Eqs.~\ref{eq_4} and~\ref{eq5}, we rewrite Eq.~\ref{eq_3} as follows:
\begin{equation}
\alpha \beta \gamma Z = \left(y \gamma - m \sec(x)\right) \left( 1 - \frac{\alpha \beta \gamma d}{\sqrt{\left(y^2 - m^2\right)\beta^2\gamma^2 - \left(y \gamma - m \sec(x)\right)^2}} \right).
\end{equation}

To simplify, we define the following variables:~$ K = y \gamma - m \sec(x), H^2 = \left(y^2 - m^2\right)\beta^2 \gamma^2 > 0, W = \alpha \beta \gamma Z$, and $G = \alpha \beta \gamma d$. This yields:
\begin{equation}
W = K \left( 1 - \frac{G}{\sqrt{H^2 - K^2}} \right).
\end{equation}

For any real value of $W$, we observe that $K < H$ is always true. Substituting $K = H \sin\phi$, where $-\pi/2 < \phi < \pi/2$, gives:
\begin{equation}\label{eq_g}
W = H \sin\phi - G \tan\phi = g(\phi).
\end{equation}

\begin{figure}[ht!]
\centering
%trim= left bottom right top
\includegraphics[trim=0cm 0cm 0cm 0cm, clip, width=6cm]{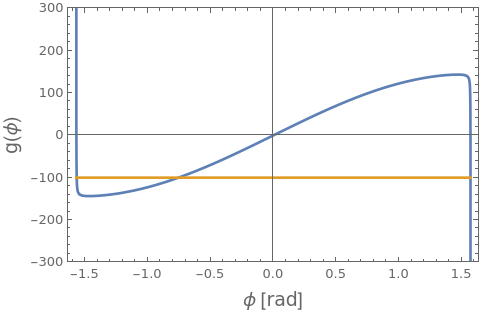}
\caption{\label{fig:g_phi} An example of the function $g(\phi)$ from Eq.~\ref{eq_g}, where $H = 145.9$, $G = 0.1645$. The orange horizontal line is the value of $W$.}
\end{figure}

The behavior of $g(\phi)$ is illustrated in Fig.~\ref{fig:g_phi}. Given that $H \gg G > 0$, we require the solution for $\phi$ to satisfy the condition that the first derivative of $g(\phi)$ is positive, i.e., $g'(\phi) > 0$. This ensures that the selected values of $\phi$ correspond to the central region of the function, where the solution is well-defined and physically meaningful.

%----------------------FIGURE 1-----------------
%----------------------------------------------

After solving the equation and determining $\phi_0$ (using Newton's method, for instance) and verifying the derivative $g'(\phi_0)$ is positive, we obtain $K = H \sin\phi_0$. Using this value, we find $\cos(x) = m / (y \gamma - K)$, and the momentum in the center-of-mass frame is given by $k = m \tan(x)$. The excitation energy $E_x$ can then be calculated using Eq.~\ref{eq_4} with the mass of the heavy recoil B ($M$) and the total invariance mass in the center-of-mass frame ($M_{cm}$):
\begin{align}
    E_x &= \sqrt{m^2 + M_{cm}^2 - 2 M_{cm} \sqrt{m^2 + k^2}} - M.
\end{align}
The center-of-mass scattering angle, $\theta_{cm}$, can be deduced using Eq.~\ref{eq5}.

%######################################################
\section{Demonstration \& Discussion}

%A program is coded to perform this task for HELIOS~\cite{Lightall2010}. The parameters $(a,b)$ are randomly picked within a certain range, usually, $1/a \in (220,320)$ and $b \in (-1, 1)$. We take the \iso{25}{Mg}($d$,$p$) reactions at 6 MeV/u with 2.85 T magnetic field to demonstrate the method (Fig.~\ref{fig:demo}). The threshold ($\tau$) was set to be 0.1 MeV, which is roughly equal to the energy resolution. The total time for the calibration process is roughly few minutes for the whole array. After the calibration, the excited energy and center-of-mass angle are calculated and shown in Fig.~\ref{fig:ex}. 

\begin{figure}[ht!]
\centering
%trim= left bottom right top
\includegraphics[trim=0cm 0cm 0cm 0cm, clip, width=13.5cm]{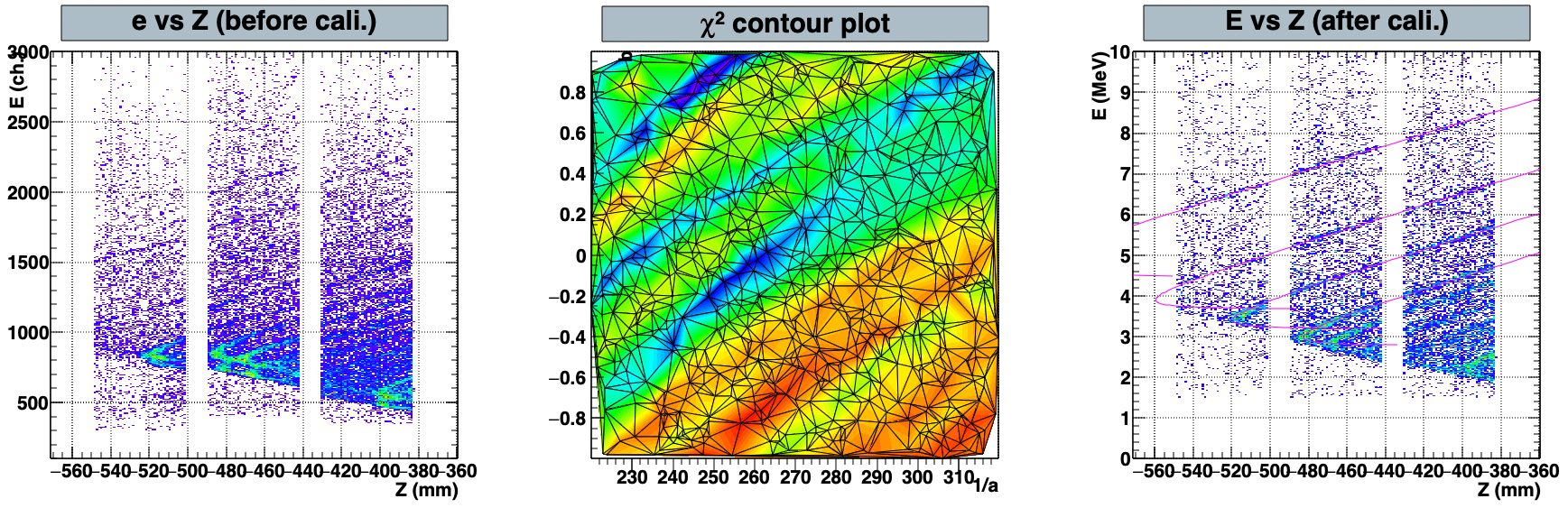}
\caption{\label{fig:demo} (Left) The measured energy \( e \) (in channels) versus measured position \( Z \) (in mm) before calibration. (Middle) A color contour plot of the \(\chi^2\) values for the fitting parameter space for the scaling factor \( a \) and offset \( b \) for a single detector. One thousand random parameter pairs (\(a, b\)) were generated, and the \(\chi^2\) and event count \(N\) were computed for each trial. Lower \(\chi^2\) values, indicating better fits, are shown in deep blue, while higher values are shown in red. (Right) The calibrated energy \( E \) (in MeV) versus measured position \( Z \). The calibrated energy \( E \) is obtained by applying the optimal linear transformation \( E = ae + b \), aligning the measured data with known physical values. The red curves represent the theoretical kinematic trajectories of the \iso{26}{Mg} states used in the energy calibration.}

\end{figure}

A program has been developed and implemented this method for HELIOS~\cite{Lightall2010}. The calibration parameters $(a, b)$ are randomly distributed within a specified range, typically $1/a \in (220, 320)$ and $b \in (-1, 1)$. To demonstrate the method, we applied it to the \iso{25}{Mg}($d$,$p$) reaction at 6 MeV/u under a magnetic field of 2.85 T (Fig.~\ref{fig:demo}).

\begin{figure}[ht!]
\centering
%trim= left bottom right top
\includegraphics[trim=0cm 0cm 0cm 0cm, clip, width=9cm]{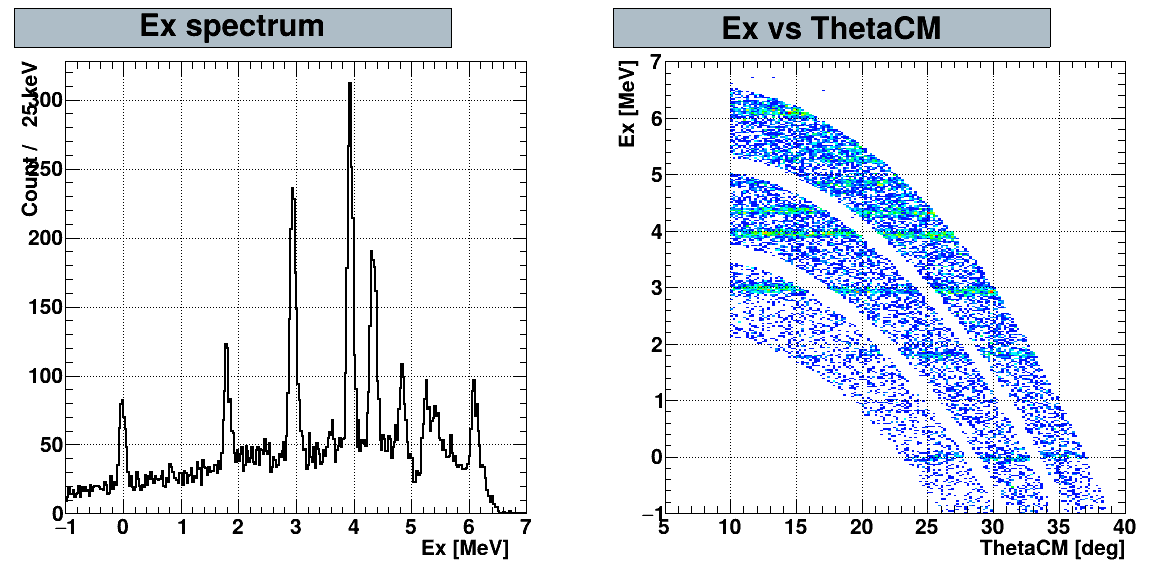}
\caption{\label{fig:ex} (Left) The excited energy spectrum of \iso{26}{Mg}. (Right) The $E_x-\theta_{cm}$ plot. A $\theta_{cm} > 10^\circ$ cut was applied.}
\end{figure}

The threshold $\tau$ was set to 0.1 MeV, approximately matching the intrinsics energy resolution of the detector. The calibration process for a single detector required only a few seconds to complete for 1000 trials, with a 3.3 GHz intel Core i5 on a 2020 iMac. Following the calibration, the excitation energy and center-of-mass angle were reconstructed and are presented in Fig.~\ref{fig:ex}, indicate that the reconstruction of the $(E_x, \theta_{cm})$ work very well.

%----------------------FIGURE 2-----------------
%----------------------------------------------

%----------------------FIGURE 3-----------------
%----------------------------------------------

In this demonstration, only the lowest four excited states (0, 1.809, 2.938, and 3.942 MeV), were used for calibration. The method correctly matched the experimental data, resulting in an accurate calibration of higher excited states (4.34 MeV and 4.84 MeV)~\cite{Burlein84}.

In the raw energy plot (left panel of Fig.~\ref{fig:demo}), four detectors are positioned at the same $z$ position. After kinematics calibration, all detectors align well with the theoretical values (right panel of Fig.~\ref{fig:demo}). The average $\chi^2/N$ is $0.03$~MeV$^2$, which means the average deviation from a valid data point to the theoretical kinematics curve is 0.17 MeV. 

No background gating was applied during the calibration, yet the results were satisfactory. Applying appropriate gates to clean the data would speed up the calibration process. In the $\chi^2$ plot (middle panel of Fig.~\ref{fig:demo}), several local minima are observed. If the number of trials is insufficient, the global minimum may not be identified. The current random parameter sampling could be optimized with a more adaptive search algorithm.

The method may fail if the level density is high, potentially leading to incorrect fitting parameters corresponding to a global minimum that matches a different set of levels. This can occur due to a larger number of data points concentrating where the level density is high, increasing the count of valid data points $N$, and result in a smaller $\chi^2$ value.

The method remains effective even with just two known states, provided that no other pair of states has a similar energy separation and the level density is not too high. This is because the slope of the $E$-$Z$ curves must match the kinematics curves. Furthermore, the range of the scaling parameter, $a$, helps to exclude spurious minima. If another pair of states has similar energy separation, the method may incorrectly fit those states, leading to incorrect calibration. In general, using more known states improves the reliability of the fit. If high-energy states are unknown, it is advisable to use a gate to select only the known states for calibration.
When only a single known state is present, and the experimental data contains a single strongly populated peak, the fitting method remains effective. This is because the fit relies on the shape of the $E-Z$ curve, which provides both the slope and energy range required for calibration. However, the accuracy for other weakly populated states cannot be guaranteed.

For the calibration method to function properly, the $z$-position of the detector array is assumed to be well known. However, if the $z$ position is not accurately measured, the characteristic bending of the $E-Z$ curve, corresponding to small $\theta_{cm}$, can be used to estimate the actual $z$ position before performing the kinematics calibration.

It is important to note that the reconstruction of $(E_x, \theta_{cm})$ is highly reliable for $\theta_{cm} \geq 10^\circ$ (for the HELIOS spectrometer). However, for $\theta_{cm} < 10^\circ$, the method does not produce satisfactory result. This is due to the intrinsic ambiguity in the inverse transformation, where a single pair of measured values $(E, Z)$ correspond to two possible solutions of $(E_x, \theta_{cm})$ when the shift of the $Z$ from $Z_0$ becomes significant (Fig.~\ref{fig:bending}). The development of a reliable method to recover data at small $\theta_{cm}$ values remains an open challenge.

%######################################################
\subsection{Compare to calibration using alpha source}

\begin{figure}[ht!]
\centering
%trim= left bottom right top
\includegraphics[trim=0cm 0cm 0cm 0cm, clip, width=9cm]{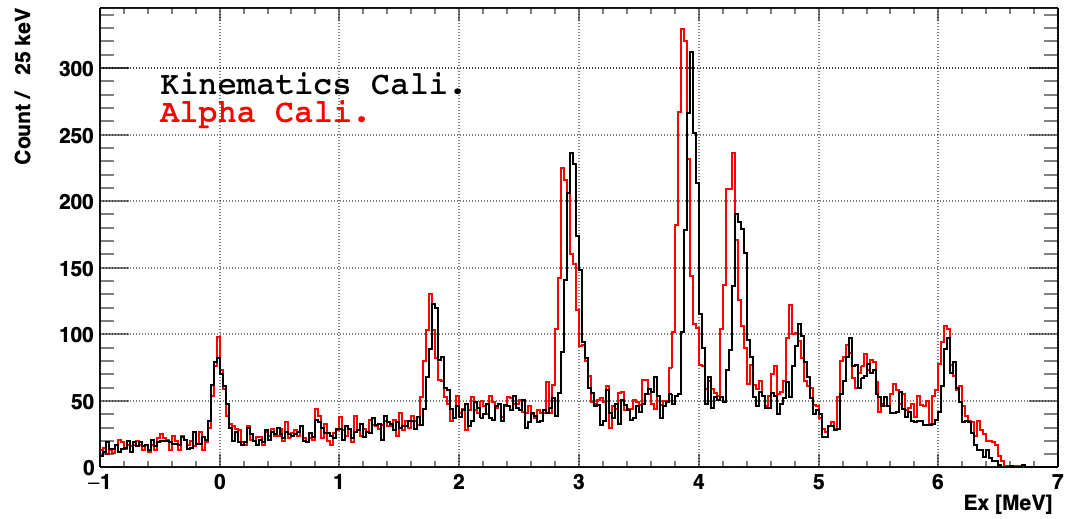}
\caption{\label{fig:ex_alpha}The comparison of the $E_x$ spectra using kinematics calibration (black) and alpha calibration (red).}
\end{figure}

Before the experiment, a mixed sealed $\alpha$ source containing $^{138}$Gd and $^{244}$Cm was used, which emit alpha particles with energies of 3.18 MeV and 5.80 MeV, respectively. Using the same reconstruction method for ($E_x$, $\theta_{cm}$) described in Section~\ref{sec_3}, the resulting $E_x$ spectrum is shown in Fig.~\ref{fig:ex_alpha}.

%----------------------FIGURE 4-----------------
%----------------------------------------------

The two spectra show slight differences. The energy resolutions ($\sigma$) for the 3.94~MeV state are 47~keV and 51~keV for the kinematics calibration (black) and alpha calibration (red), respectively. However, the peak corresponding to the 3.94~MeV state (black) is shifted to 3.87~MeV (red) when using alpha calibration. This slight inaccuracy in the $E_x$ spectrum with alpha calibration arises from the limited energy range of the alpha source, whereas the kinematics calibration utilizes the full energy range of the light-recoil particles. Nevertheless, even when using only alpha calibration, the reconstruction of $E_x$ and $\theta_{cm}$ via the inverse transform described in Section~\ref{sec_3} remains acceptable.

%######################################################
%\subsection{Recovery of small angle data}

%The detection position $Z$ is different from the position $Z_0$ due to the finite detector size (Fig.~\ref{fig:bending}). The effect on the $E-Z$ plot is shown in Fig.~\ref{fig:finite_detector}. At very small $\theta_{cm}$, the transformation $(E, Z) \rightarrow (E_x, \theta_{cm})$ is not unique (or not bijective), i.e. there are two solutions of $(E_x, \theta_{cm})$ for a give $(E,Z)$,  which can render the experimental data unusable. However, when the level spacing is sufficiently large such that the $E_Z$ curve of a particular state does not overlap with the neighbor states, Eq.~\ref{eq_3} can be used to deduce $\theta_{cm}$ for a given $E_x$.

%----------------------FIGURE 4-----------------
%\begin{figure}[ht!]
%\centering
%trim= left bottom right top
%\includegraphics[trim=0cm 0cm 0cm 0cm, clip, width=9cm]{Fig5.png}
%\caption{\label{fig:finite_detector}The planes of $(E_x,\theta_{cm})$ of Eq~\ref{eq1} (orange, infinite small detector) and Eq.~\ref{eq2} (blue, finite size detector) on $(E, Z)$ coordinate. The range of $\theta_{cm}$ is from 4$^\circ$ to $30^\circ$, and the range of the $E_x$ is from 0 to 3 MeV.}
%\end{figure}
%----------------------------------------------

%For example, the 2.9 MeV state of $^{26}Mg$ is separated from the 1.8 MeV state and the 3.9 MeV state, by gating on the $E-Z$ plot and isolate 

%######################################################
\section{Summary}

A novel method is presented for obtaining the excited energy spectrum and angular distribution from solenoidal spectrometers used in nuclear reaction studies. These spectrometers measure the energy ($e$) and position ($Z$) of the light recoil particles, which are related to the excitation energy ($E_x$) of the heavy recoil nucleus and center-of-mass scattering angle ($\theta_{cm}$). Conventional methods, which rely on projecting $E-Z$ curves, face challenges at forward angles due to detector geometry effects. Our new approach addresses these limitations by first calibrating the experimental $e-Z$ data (in channel - mm) to obtain $E-Z$ values (in MeV - mm) using known excited states and a minimum chi-squared fitting procedure with a distance threshold to reject noise. Then, through a series of transformations based on relativistic kinematics and cyclotron motion in the spectrometer's magnetic field, we derive an analytical relationship that enables a direct inverse transformation from the calibrated $E-Z$ data to $E_x$ and $\theta_{cm}$ simultaneously. This method circumvents the non-linearity of the $E$-$Z$ relationship and ensures consistent treatment of all detectors. The efficacy of this method is demonstrated by applying it to the \iso{25}{Mg}($d$,$p$) reaction. This method is already automated in HELIOS, provides a robust, speedy, and accurate way to extract excitation energy spectrum and angular distributions.

%######################################################
\section{Acknowledgement}

This research utilized resources from Florida State University's John D. Fox Laboratory, supported by the National Science Foundation under Grant No. PHY-2412808, and Argonne National Laboratory’s ATLAS facility, a Department of Energy Office of Science User Facility supported by the U.S. Department of Energy, Office of Science, Office of Nuclear Physics, under Contract No. DE-AC02-06CH11357.

%\bibliographystyle{elsarticle-num} 
%\bibliography{cas-refs}

\begin{thebibliography}{00}



\bibitem{Wuosmaa2007} A.H.~Wuosmaa {\it et al.}, \href{https://doi.org/10.1016/j.nima.2007.07.029}{Nuclear Instruments and Method A {\bf 580}, 1290-1300 (2007)}

\bibitem{Tang2023} T.L.~Tang, \href{https://wiki.anl.gov/heliosdaq/File:Kinematics_of_HELIOS.pdf}{Kenematics of HELIOS.pdf}

\bibitem{Lightall2010} J.C.~Lightall {\it et al.}, \href{https://doi.org/10.1016/j.nima.2010.06.220}{Nuclear Instruments and Method A {\bf 622}, 97-106 (2010)}


\bibitem{Burlein84} M.~Burlein, K.~S.~Dhuga, and H.~T.~Fortune, \href{https://doi.org/10.1103/PhysRevC.29.2013}{Phys. Rev. C {\bf 29} (2013)}

\end{thebibliography}


\end{document}